\documentclass[aps,prc,amsmath,preprint,superscriptaddress,showpacs]{revtex4}

\newcommand{\bra}[1]{\langle #1 |}
\newcommand{\ket}[1]{| #1 \rangle}
\usepackage{bm,graphicx}

\begin{document}

%\preprint{}

%Title of paper
\title{
Unified Analysis of Spin Isospin Responses of Nuclei}

%\thanks{}
%\altaffiliation{}
\author{T.~Wakasa}
\affiliation{Department of Physics, Kyushu University,
Higashi, Fukuoka 812-8581, Japan}
\author{M.~Ichimura}
\affiliation{Faculty of Computer and Information Sciences, Hosei University, 
Koganei, Tokyo 184-8584, Japan}
\author{H.~Sakai}
\affiliation{Department of Physics, The University of Tokyo, 
Bunkyo, Tokyo 113-0033, Japan}

\date{\today}

\begin{abstract}
% insert abstract here
We investigated the Gamow-Teller (GT) strength distribution, 
especially the quenching with respect to the GT sum rule, 
and the enhancement of the pionic responses 
in the quasielasic scattering region, 
in the same theoretical framework. 
That is the continuum random phase approximation 
with the $\pi+\rho+g'$ model interaction, 
incorporated with distorted wave impulse approximation 
and two-step calculations. 
 From this analysis we searched the Landau-Migdal parameters, 
$g'_{NN}$ and $g'_{N\Delta}$, 
through the comparison with the experimental data 
of the GT strength distribution obtained at 300 MeV 
and the spin-longitudinal (pionic) cross sections 
$ID_q$ of $(\vec{p}, \vec{n})$ at 350 and 500 MeV. 
This comprehensive and sophisticated study gave 
a common set of $g'_{NN}$=0.6--0.7 and $g'_{N\Delta}$=0.2--0.4, 
for both low and high momentum transfers. 
\end{abstract}

% insert suggested PACS numbers in braces on next line

%\maketitle must follow title, authors, abstract, \pacs, and \keywords
\maketitle

Recent $(p,n)$ and $(n,p)$ experiments at intermediate energies 
have presented reliable information 
on nuclear spin isospin responses \cite{rapaport_alfold}. 
Following two contrastive subjects are especially interesting.  
One is quenching of the total strength 
of the Gamow-Teller (GT) transitions \cite{npa_396_127c_1983}
from its sum rule value $3(N-Z)$ \cite{npa_334_248_1980}, 
and the other is enhancement of the pionic (isovector spin-longitudinal) 
response functions in the quasielastic scattering (QES) region 
\cite{prl_73_3516_1994,prc_59_3177_1999,prc_69_054609_2004} 
as a precursor of the pion condensation \cite{np_a379_429_1982}. 
A common key concept to understand these contrastive phenomena simultaneously 
is the Landau-Migdal (LM) parameters, 
$g'_{NN}$, $g'_{N\Delta}$, and $g'_{\Delta\Delta}$, 
which specify the LM interactions $V_{\rm LM}$, namely 
the zero range particle-hole ($ph$) and Delta-hole ($\Delta h$) interactions.

In this paper we study the GT strength distributions 
and the quenching factor observed at 295 MeV at RCNP, 
and the spin-longitudinal cross sections 
$ID_q$ of $(\vec{p}, \vec{n})$ at 346 MeV at RCNP and at 494 MeV at LAMPF, 
which well represent the pionic response function $R_{L}$. 
To all these observables, we apply the same theoretical framework, 
namely the continuum random phase approximation (RPA) 
with the $\pi+\rho+g'$ model interaction, 
which properly treats the finite geometry. Then we look for 
the LM parameters, which reproduce the data as well as possible. 

Estimations of $g'_{NN}$ from the GT giant resonances (GTGR) have been 
carried out by several authors \cite{rmp_64_491_1992}. 
For instance, Suzuki \cite{plb_104_92_1981}
used energy weighted sum technique 
and Bertsch, Cha, and Toki \cite{prc_24_533_1981} used the continuum RPA. 
Fitting the peak position of GTGR, 
they obtained similar values of $g'_{NN}\approx$0.6 for $^{90}$Zr.
Their analyses used only the LM interaction for the nucleons. 
Most of later works with $\Delta$ \cite{npa_395_189_1983,prc_31_372_1985}
used the universality ansatz, $g'_{NN} = g'_{N\Delta} = g'_{\Delta\Delta}$. 
Further the experimental spectra are of the cross sections of $(p, n)$, 
but not of the genuine GT strength. 
We re-investigate the GTGR spectrum 
by use of $\pi+\rho+g'$ model interaction 
without the universality ansatz.

 From the GT quenching factor, 
Suzuki and Sakai \cite{plb_455_25_1999} 
estimated $g'_{N\Delta}$ by the Fermi gas model only with $V_{\rm LM}$, 
treating the finite size effect crudely. 
They obtained $g'_{N\Delta} \approx 0.2$ for $^{90}$Zr. 
Using the first order perturbation on the $N\Delta$ transition part of 
the $\pi + \rho + g'$ model interaction, 
Arima {\it et al.} \cite{plb_499_104_2001} 
obtained $g'_{N\Delta} \approx 0.3$.  
This increase of 0.1 from the Suzuki-Sakai's result 
comes from the $\pi$ and $\rho$ exchange interaction 
due to the nuclear finite size. 
The present RPA calculation presents more integrated analysis.

As to the pionic responses in the QES region, it has been 
shown \cite{prc_63_044609_2001}
that a conventionally used eikonal approximation is 
not quantitatively reliable 
to extract the pionic response function $R_{L}$ from $ID_q$ 
due to the nuclear distortion. 
Thus we have to calculated $ID_q$ 
by the distorted wave impulse approximation (DWIA) 
incorporated with the continuum RPA response functions, 
and to compare the theoretical and experimental results directly. 
Further it has also been shown \cite{prc_69_054609_2004}
that two-step processes contribute as an appreciable background 
for the RCNP data. Therefore we extend the DWIA + two-step analysis 
to the LAMPF data and search the suitable $g'$'s in an overall point of view.

 We write the $\beta^{\pm} ({\rm GT}^{\pm})$  
transition operators with {\it N} and $\Delta$ in the unit of $g_A$ as 
\begin{equation}
O^{\pm}_{\rm GT} = \mp\frac{1}{\sqrt{2}}\sum_{k=1}^{A} \left\{
  \tau_{k,\pm 1}\bm{\sigma}_{k} 
+ \frac{g_A^{N\Delta}}{g_A}\left( 
T_{k,\pm 1}\bm{S}_{k}-T^{\dagger}_{k,\mp 1}\bm{S}^{\dagger}_{k} \right)
\right\},
\end{equation}
with $\tau_{\pm 1}=\mp \frac{1}{\sqrt{2}}(\tau_x \pm i\tau_y)$ and 
$T_{\pm 1}=\mp\frac{1}{\sqrt{2}}(T_x \pm i T_y)$,
where $g_A (g_A^{N\Delta})$ is the weak coupling constant 
for {\it NN} ($N\Delta$) transition, 
and $\bm{\sigma} (\bm{\tau})$ is the nucleon Pauli spin (isospin) matrix, 
and $\bm{S} (\bm{T})$ is the spin (isospin) transition operator 
from {\it N} to $\Delta$. 
 Similarly we write the isovector spin-longitudinal transition operators 
with momentum transfer $\bm q$ as 
\begin{equation}
 O^{\lambda}_{L}(\bm{q}) = \sum_{k=1}^{A} \left\{
\tau_{k,\lambda}\bm{\sigma}_{k}\cdot\hat{\bm{q}} 
+ \frac{f_{\pi N\Delta}}{f_{\pi NN}}\left( 
T_{k,\lambda}\bm{S}_{k}\cdot\hat{\bm{q}} +(-1)^{\lambda}
T^{\dagger}_{k,-\lambda}\bm{S}^{\dagger}_{k}\cdot\hat{\bm{q}} \right)
\right\}e^{i\bm{q}\cdot\bm{r}_{k}}\ ,
\end{equation}
where $\lambda = 0,\  \pm 1$, and 
$f_{\pi NN} (f_{\pi N\Delta})$ is the $\pi NN (\pi N\Delta)$ 
coupling constant. 
Here we neglect the transitions from $\Delta$ to $\Delta$ 
in both $O^{\pm}_{\rm GT}$ and $O^{\lambda}_{L}(\bm{q})$.
We take the quark model relation 
$f_{\pi N \Delta}/f_{\pi NN}$ = 
$g^{N\Delta}_A/g_A$ =
$\sqrt{72/25}$ $\simeq $ 1.70 in this paper. 
What we are interested in are how the nuclei respond to these operators.

Since neither the momentum $\bm{q}$ nor 
the spin directions are conserved in finite nuclei,
we introduce the spin-isospin transition densities 
\begin{equation}
O^{N}_{\lambda,a}(\bm{r}) = 
\sum_{k=1}^{A} \tau_{k,\lambda}\sigma_{k,a}\delta(\bm{r}-\bm{r}_k)\ , \quad 
O^{\Delta}_{\lambda,a}(\bm{r}) = 
\sum_{k=1}^{A} T_{k,\lambda}S_{k,a}\delta(\bm{r}-\bm{r}_k)\ ,
\end{equation}
with $a=x, y, z$, and calculate the spin-isospin response functions 
\begin{eqnarray}
R^{\alpha\beta}_{\lambda,ba}(\bm{r}',\bm{r},\omega) 
= \sum_{n}
\bra{\Psi_0}O^{\alpha\dagger}_{\lambda,b}(\bm{r}')\ket{\Psi_n}
\bra{\Psi_n}O^{\beta        }_{\lambda,a}(\bm{r} )\ket{\Psi_0} 
\delta(\omega-({\cal E}_{n}-{\cal E}_{0}))\ ,
\end{eqnarray}
by the continuum RPA with the orthogonality condition 
in the coordinate representation \cite{prc_51_269_1995}.

The $\pi$+$\rho$+$g'$ model interaction is written as 
\begin{equation}
V^{\rm eff}(\bm{q},\omega)
= V_{\rm LM}+V_{\pi}(\bm{q},\omega) +  V_{\rho}(\bm{q},\omega)\ ,
\label{eq:Veff}
\end{equation}
where $V_{\pi}$ and $V_{\rho}$ are the one-pion 
and the one-rho-meson exchange interactions, respectively. 
 The LM interaction $V_{\rm LM}$ is written by the LM parameters 
 $g'_{NN}$, $g'_{N\Delta}$, $g'_{\Delta\Delta}$ as 
\begin{eqnarray}
V_{\rm LM} 
&=& \frac{f_{\pi NN}^2}{m_{\pi}^2}g'_{NN} 
(\bm{\tau}_1\cdot\bm{\tau}_2)(\bm{\sigma}_1\cdot\bm{\sigma}_2) \nonumber \\
&+& \frac{f_{\pi NN}f_{\pi N\Delta}}{m_{\pi}^2}g'_{N\Delta} 
\left\{
(\bm{\tau}_1\cdot\bm{T}_2)       (\bm{\sigma}_1\cdot\bm{S}_2) + 
(\bm{\tau}_1\cdot\bm{T}^{\dag}_2)(\bm{\sigma}_1\cdot\bm{S}^{\dag}_2)
    + (1\leftrightarrow 2)
\right\} \nonumber \\
&+& \frac{f_{\pi N\Delta}^2}{m_{\pi}^2}g'_{\Delta\Delta}
\left\{(\bm{T}_1\cdot\bm{T}^{\dag}_2)(\bm{S}_1\cdot\bm{S}^{\dag}_2) 
+ (1\leftrightarrow 2) \right\}\ .
\end{eqnarray}
We fixed $g'_{\Delta\Delta}=0.5$ \cite{prc_23_1154_1981} 
since the calculated results depend on it very weakly .
Non-locality of mean fields is taken into account 
by a local effective mass approximation in the form of 
\begin{equation}
m^*(r) = m_N-\frac{f_{\rm WS}(r)}{f_{\rm WS}(0)}(m_N-m^*(0))\ ,
\label{eq:effmass}
\end{equation}
with the Woods-Saxon radial form $f_{\rm WS}(r)$.

First we discuss the strength distributions of the ${\rm GT}^-$ transitions, 
which are expressed by the ${\rm GT}^\pm$ response functions 
for the ground-state $\ket{\Psi_0}$ 
\begin{eqnarray}
R_{\rm GT}^\pm(\omega) = 
\sum_{n\ne 0}|\bra{\Psi_n}O^{\pm}_{\rm GT}\ket{\Psi_0} |^2
\delta(\omega-({\cal E}_{n}-{\cal E}_{0})),   \label{eq:RGT}
\end{eqnarray}
where $\ket{\Psi_n}$ and ${\cal E}_n$ denote 
the $n$-th nuclear state and its energy.
They are experimentally extracted 
from the $\Delta J^{\pi}=1^+$ cross sections 
$d^2\sigma_{1^+}(q,\omega)/d\Omega d\omega$ 
deduced by the multipole decomposition analysis (MDA) as 
\cite{prc_55_2909_1997}
\begin{equation}
\frac{d^2\sigma_{1^+}(q,\omega)}{d\Omega d\omega}
=
{\hat{\sigma}_{\rm GT}}F(q,\omega)R_{\rm GT}^\pm(\omega)\ ,
\label{eq:sigma_l0_vs_rgt}
\end{equation}
with the GT unit cross section ${\hat{\sigma}_{\rm GT}}$ and
the $(q,\omega)$ dependence factor $F(q,\omega)$. 

Converting the calculated response functions 
$R^{\alpha\beta}(\bm{r}',\bm{r},\omega)$ into 
those in the momentum representation $R^{\alpha\beta}(\bm{q}',\bm{q},\omega)$, 
the GT response functions $R_{\rm GT}^\pm(\omega)$ of Eq.~(\ref{eq:RGT}) 
are given by 
\begin{equation}
R_{\rm GT}^\pm(\omega)
= 
\frac{1}{2}\sum_{a}
\left\{
 R_{\pm 1,aa}^{NN}(\omega) 
+2\frac{g^{N\Delta}_A}{g_A}R_{\pm 1,aa}^{N\Delta}(\omega)
+\left(\frac{g^{N\Delta}_A}{g_A}\right)^2
R_{\pm 1,aa}^{\Delta\Delta}(\omega)
\right\}\ ,
\end{equation}
where 
$R^{\alpha\beta}(\omega)$ = 
$R^{\alpha\beta}(\bm{q}'=0,\bm{q}=0,\omega)$.
The strength distribution $R_{\rm GT}^{-}(\omega)$ 
from ${}^{90}{\rm Zr}$ to ${}^{90}{\rm Nb}$ was obtained 
by MDA of the $(p,n)$ data 
\cite{prc_55_2909_1997,jpsj_73_1611_2004}, which 
cover not only the GTGR region but also the excitation energy up to 50 MeV.

\begin{figure}
\begin{center}
\includegraphics[width=0.8\linewidth,clip]{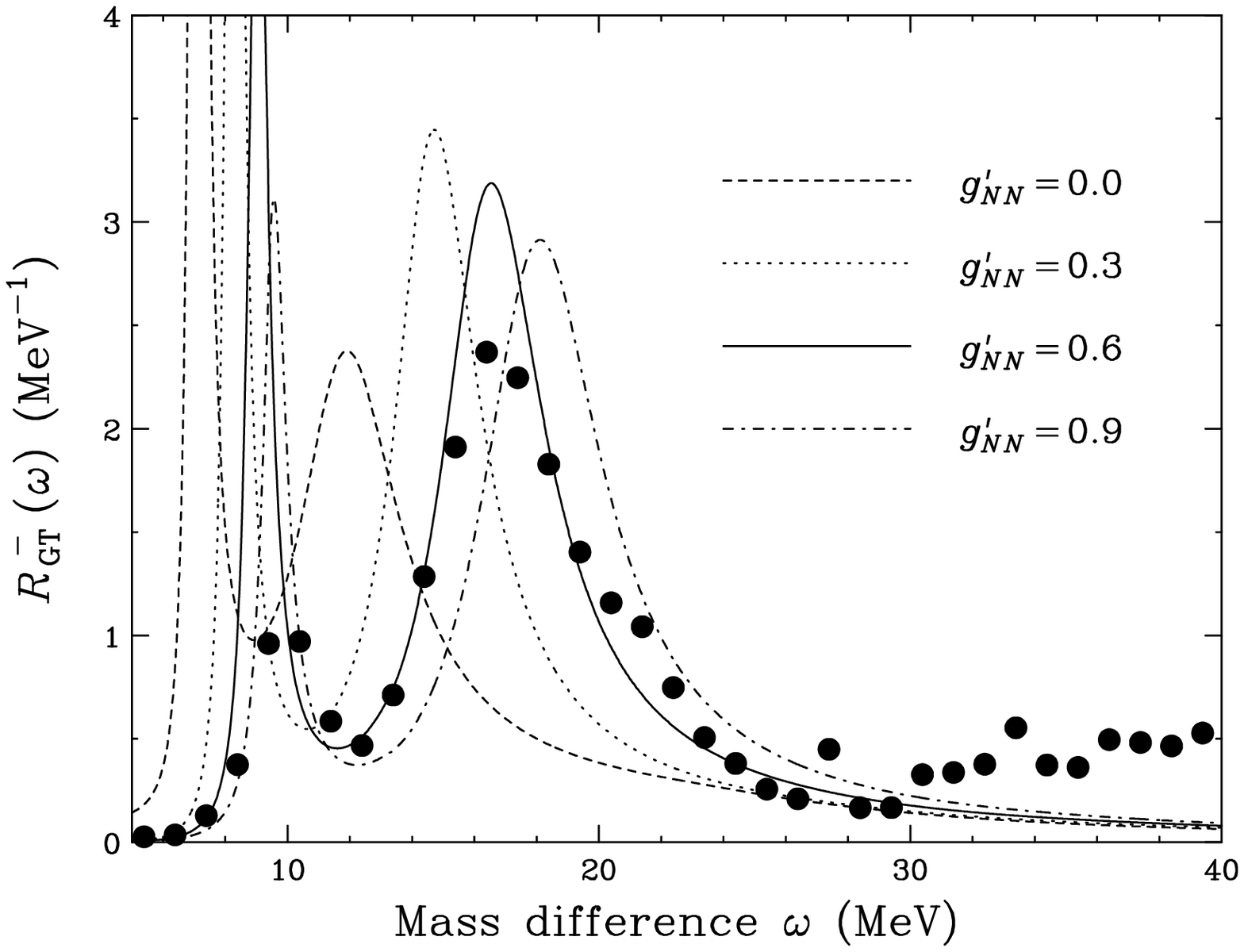}
\end{center}
\caption{The $g'_{NN}$ dependence of 
${\rm GT}^-$ strength distributions from 
${}^{90}{\rm Zr}$ to ${}^{90}{\rm Nb}$, where 
$g'_{N\Delta}$ and $m^*(0)/m_N$ are set to 0.3 and 0.7, respectively.
The filled circles are the experimental data taken from 
Ref.~\cite{prc_55_2909_1997}.}
\label{fig1}
\end{figure}

 Figure~\ref{fig1} shows the $g'_{NN}$ dependence of the GTGR peak position. 
 The curves correspond to the results of $g'_{NN}$=0.0--0.9 in 0.3 steps, 
with $g'_{N\Delta} = 0.3$ and $m^*(0)/m_N = 0.7$.
 The result with $g'_{NN}$=0.6 reproduces the peak position well. 
This value is very close to 
the previous results \cite{plb_104_92_1981,prc_24_533_1981}. 
The excess of the theoretical value around the peaks should be redistributed
by the configuration mixing of $2p2h$ states and more \cite{prc_26_1323_1982}.
This redistribution is seen as significant experimental strength beyond GTGR.
This is the quenching of one kind, which should be distinguished from
the other quenching due to $\Delta h$ mixing discussed below.

\begin{figure}
\begin{center}
\includegraphics[width=0.8\linewidth,clip]{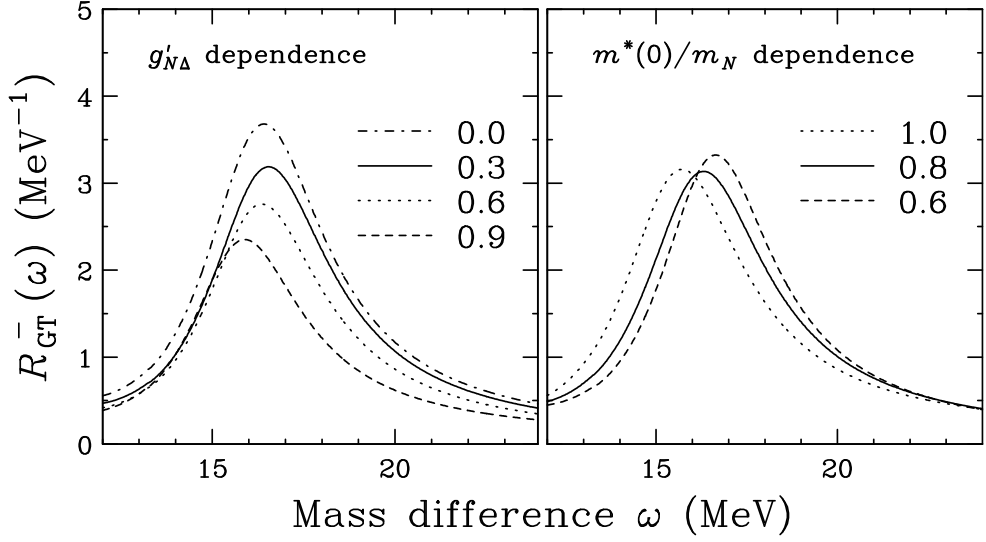}
\end{center}
\caption{
The $g'_{N\Delta}$ (left panel) and $m^*(0)/m_N$ (right panel) 
dependences of the RPA calculations.
In the left panel 
$g'_{NN}$ and $m^*(0)/m_N$ are set to 0.6 and 0.7, respectively.
In the right panel 
$g'_{NN}$ and $'_{N\Delta}$ are set to 0.6 and 0.3, respectively.}
\label{fig2}
\end{figure}

Figure~\ref{fig2} examines the $g'_{N\Delta}$ and $m^*$ dependences 
of the GTGR spectrum. 
In the left panel the curves denote the results of 
$g'_{N\Delta}$=0.0--0.9 in 0.3 steps.
The peak position hardly depends on $g'_{N\Delta}$ 
but its height strongly does.
Since the $g'_{N\Delta}$ governs the coupling between $ph$ and $\Delta h$, 
it controls how much the ${\rm GT}^{-}$ strength in the GTGR region moves up
to the $\Delta h$ region.
 The $m^*$ dependence is shown in the right panel, 
where the curves represent the results of $m^*(0)/m_N$=1.0--0.6 in 0.2 steps.
 The $m^*$ affects the GTGR so weakly that 
it is hard to get information about $m^*$.
 From these analysis, we obtained $g'_{NN}=0.6\pm 0.1$ 
with care for the small $g'_{N\Delta}$ and $m^*$ dependences.

Next we discuss the GT quenching factor $Q$, which is defined as
\begin{equation}
Q = \frac{S_{\rm GT}^-(\omega_{\rm top}^-)-S_{\rm GT}^+(\omega_{\rm top}^+)}
{3(N-Z)}\ ,
\end{equation}
with
\begin{equation}
S_{\rm GT}^\pm(\omega_{\rm top}^\pm) = 
\int^{\omega_{\rm top}^\pm}R_{\rm GT}^\pm(\omega) d\omega\ .
\end{equation}
Very recently Yako {\it et al.} \cite{nucl_ex_0411011} applied MDA to both 
the ${}^{90}{\rm Zr}(p,n)$ and ${}^{90}{\rm Zr}(n,p)$ data, 
and obtained $Q=0.88 \pm 0.03$ 
where they used the end energy $\omega_{\rm top}^- = 57$ MeV 
(50 MeV of ${}^{90}{\rm Nb}$ excitation) and chose 
the relevant $\omega_{\rm top}^+$ in account of the Coulomb energy shift. 

\begin{figure}
\begin{center}
\includegraphics[width=0.8\linewidth,clip]{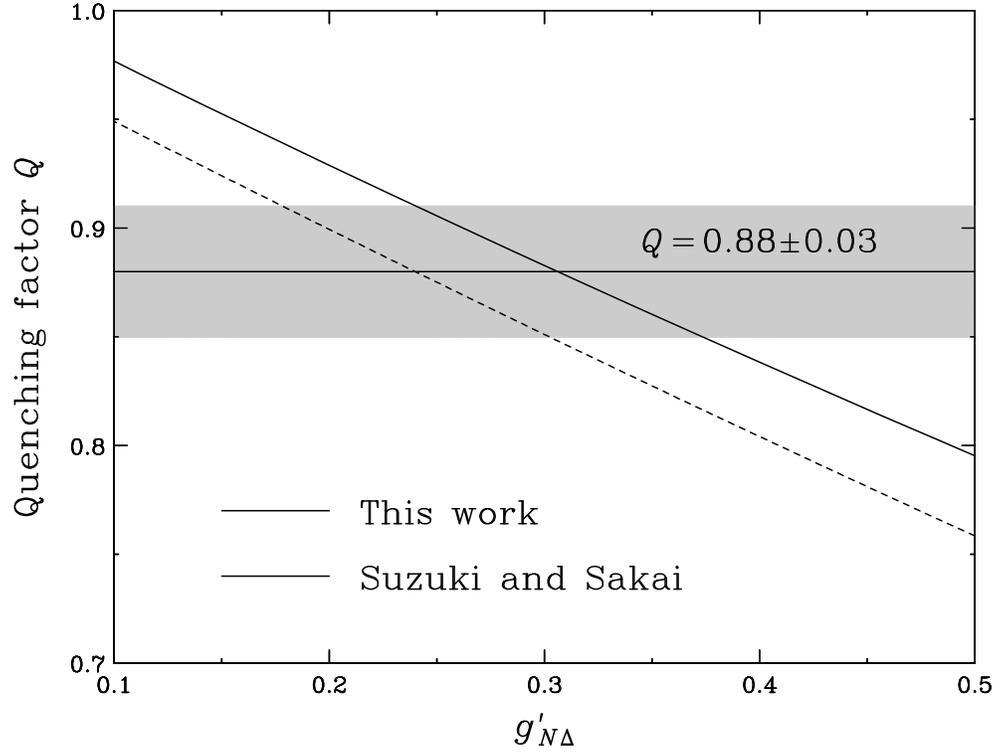}
\end{center}
\caption{
 The GT quenching factor $Q$ as a function of $g'_{N\Delta}$.
 The experimental result of $0.88\pm 0.03$ \cite{nucl_ex_0411011}
is shown by the horizontal solid line and band.
 The dashed curve is the theoretical prediction by 
Suzuki and Sakai \protect{\cite{plb_455_25_1999}}.}
\label{fig3}
\end{figure}

Since $Q$ is almost exclusively determined by $g'_{N\Delta}$ 
in the calculation, we display the $g'_{N\Delta}$ dependence 
in Fig.~\ref{fig3} with fixed $g'_{NN}=0.6$. The solid line shows 
the results of the continuum RPA and the dashed line does those of 
Suzuki-Sakai's formulas \cite{plb_455_25_1999}. 
The experimental $Q$ and its uncertainty are shown by 
the horizontal solid line and the horizontal band, respectively. 
 From this comparison we obtained $g'_{N\Delta} = 0.31\pm 0.07$. 
The difference between the present calculation and the Suzuki-Sakai's line 
is well understood by the mechanism of Arima {\it et al.} 
\cite{plb_499_104_2001}. 

Third we investigate the enhancement of the pionic modes in the QES region.
The relevant spin-longitudinal cross sections $ID_q$ 
were measured for ${}^{12}{\rm C}$ and ${}^{40}{\rm Ca}$ at $T_p$=346 MeV 
\cite{prc_59_3177_1999,prc_69_054609_2004} and 494 MeV 
\cite{prl_73_3516_1994} taken at RCNP and LAMPF, respectively. 
We performed DWIA calculations by use of 
the response functions $R^{\alpha\beta}(\bm{r}',\bm{r},\omega)$, 
and also estimated the two-step contributions 
in the same manner as Ref.~\cite{prc_69_054609_2004}. 

\begin{figure}
\begin{center}
\includegraphics[width=0.7\linewidth,clip]{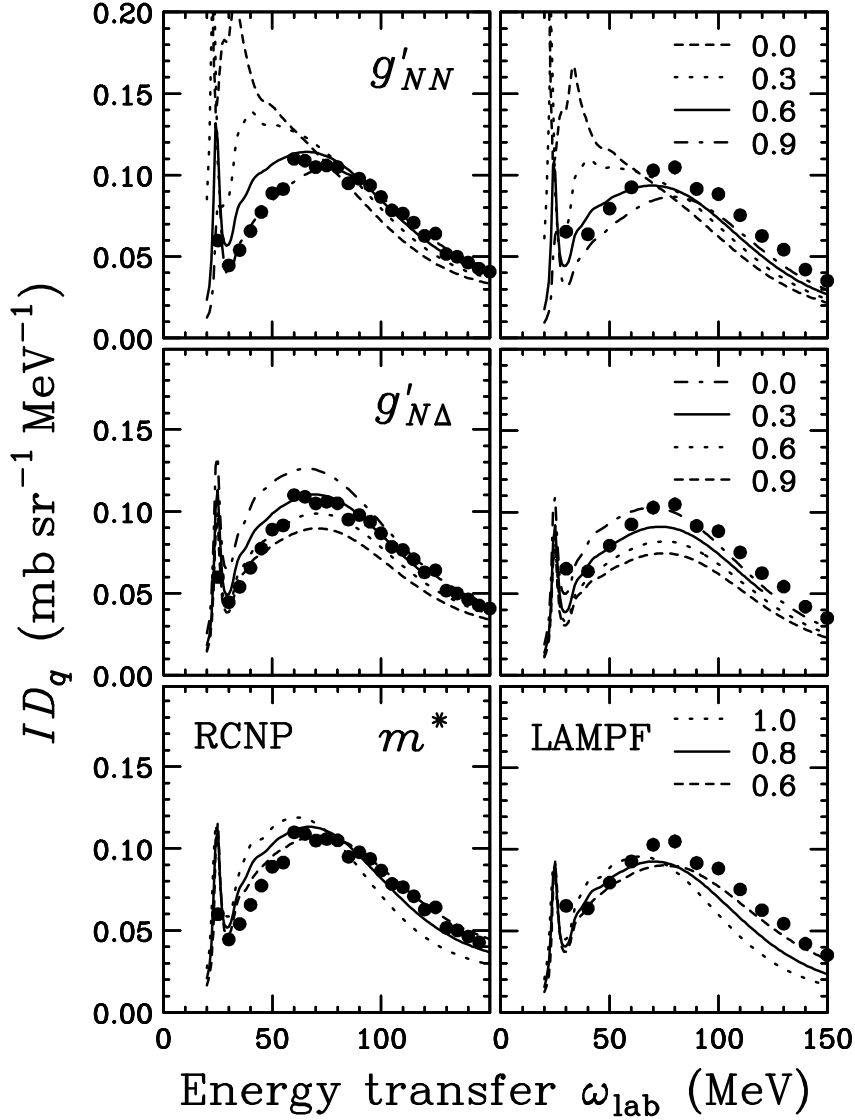}
\end{center}
\caption{
 The spin-longitudinal polarized cross section $ID_q$ 
for the ${}^{12}{\rm C}$ reaction at 
$T_p$=346 MeV \cite{prc_59_3177_1999,prc_69_054609_2004} (left panels) 
and $T_p$=494 MeV \cite{prl_73_3516_1994} (right panels).
 The top, middle, and bottom panels show the 
$g'_{NN}$, $g'_{N\Delta}$, and $m^*(0)/m_N$ 
dependences of the calculations.
 The notations of the curves are the same as those 
in Figs.~\protect{\ref{fig1}} and \protect{\ref{fig2}} 
except for $g'_{NN}$=0.7 in the middle and bottom panels.}
\label{fig4}
\end{figure}

Since the obtained characteristics are very similar 
both for ${}^{12}{\rm C}$ and ${}^{40}{\rm Ca}$, 
we compare in Fig.~\ref{fig4} the calculations with 
the experimental $ID_q$ only for ${}^{12}{\rm C}$ 
taken at RCNP and LAMPF in the left and right panels, respectively.

The top panels show the $g'_{NN}$ dependence for $g'_{NN}$=0.0--0.9 
in 0.3 steps with the fixed $g'_{N\Delta}=0.3$ and $m^*(0)/m_N = 0.7$. 
The calculations are sensitive to $g'_{NN}$ near and below the QES peak.
The experimental data are reasonably reproduced with 
$g'_{NN}$=0.7, whose uncertainty could be about 0.1.
This result is consistent with the value of $g'_{NN}=0.6\pm 0.1$ 
evaluated from the GTGR spectrum. 

The middle panels denote the $g'_{N\Delta}$ dependence for 
$g'_{N\Delta}$=0.0--0.9 in 0.3 steps with the fixed $g'_{NN} = 0.7$ 
and $m^*(0)/m_N = 0.7$. 
The dependence is evidently seen around the QES peak.
 The most probable choices of $g'_{N\Delta}$ are about 
0.4 and 0.2 for the RCNP and LAMPF data, respectively.
 From these results, we could estimate $g'_{N\Delta}= 0.3\pm 0.1$, 
which is consistent with that from the quenching factor $Q$.

 The bottom panels display the $m^*$ dependence for 
$m^*(0)/m_N$=1.0--0.6 in 0.2 steps with the fixed $g'_{NN} = 0.7$ and 
$g'_{N\Delta} = 0.3$. 
The theoretical estimations \cite{pl_126b_421_1983,np_a481_381_1988} of 
$m^*(0)/m_N \approx 0.7$ is consistent with that required by the data. 

In summary, we reported the theoretical analyses 
of the two contrastive phenomena, the quenching of the GT transition and 
the enhancement of the pionic response at the QES region.
The GT strength distributions and the latest data of the quenching factor 
are calculated by means of the continuum RPA 
with the $\pi + \rho + g'$ interactions including $\Delta$ degree of freedom. 
In the same framework of the structure calculation 
incorporated with the DWIA and two-step calculations, we also calculated 
the spin-longitudinal cross sections $ID_p$ at different incident energies. 
By this elaborated and comprehensive calculations 
in a unified way for widespread phenomena, 
we found a common set of the LM parameters, 
$g'_{NN}$=0.6--0.7 and $g'_{N\Delta}$=0.2--0.4, 
which explain the quenching and the enhancement simultaneously. 

% If you have acknowledgments, this puts in the proper section head.
%\begin{acknowledgments}
% put your acknowledgments here.
 We thank all the members of the RCNP-E57, E59, E131, and E149 
collaborations, especially for K. Yako.
 This work was supported in part by the 
Grands-in-Aid for Scientific Research Nos. 
12640294, 12740151, and 14702005 of the 
Ministry of Education, Culture, Sports, 
Science, and Technology of Japan.
%\end{acknowledgments}

%
%References
%

\clearpage

\clearpage

\clearpage

\clearpage

\end{document}